\documentclass[11pt,preprintnumbers,aps,amssymb,nofootinbib,amsmath,superscriptaddress]
{revtex4}
\usepackage{epsfig,epsf}
\usepackage{bm} 
\usepackage{slashed}
\usepackage{color} 
\newcommand{\beq}{\begin{equation}}
\newcommand{\beql}[1]{\begin{equation}\label{#1}}
\newcommand{\eeq}{\end{equation}}
\def\bal#1\gal{\begin{align}#1\end{align}}
\newcommand{\ball}[1]{\bal\label{#1}}
\newcommand{\eq}[1]{(\ref{#1})}
\newcommand{\fig}[1]{Fig.~\ref{#1}}
\renewcommand{\sec}[1]{Sec.~\ref{#1}}
\newcounter{topiccounter}
\renewcommand{\b}[1]{{\bm #1}} 
\newcommand{\unit}[1]{\hat {{\bm #1}}} 

\newcommand{\Tr}{{\,\rm Tr}}
\newcommand{\im}{\,\mathrm{Im}\!}

\newcommand{\e}{\varepsilon}
\newcommand{\aver}[1]{\left\langle #1 \right\rangle}

\begin{document}

\title{Excitation of Chandrasekhar-Kendall photons in  Quark Gluon Plasma\\ by ultrarelativitsic quark}

\author{Kirill Tuchin}

\affiliation{Department of Physics and Astronomy, Iowa State University, Ames, Iowa, 50011, USA}

\date{\today}


\begin{abstract}

A quark propagating through the quark gluon plasma and scattering off the thermal gluons can radiate photons in states with definite angular momentum and magnetic helicity. These states, known as the Chandrasekhar-Kendall states, are eigenstates of the curl operator and have a non-trivial topology.  I compute the spectrum of these states in the ultrarelativistic limit and study its properties.

\end{abstract}

\maketitle

\section{Introduction}\label{sec:a}

The electromagnetic radiation is a precise tool to study the dynamics of the quark-gluon plasma (QGP) \cite{Shuryak:1978ij,Hwa:1985xg,Kapusta:1991qp,Baier:1991em,Aurenche:1998gv,Arnold:2001ba,Arnold:2002ja,Bass:2004de,Turbide:2007mi}. Significant process has been made over the recent decade \cite{vanHees:2011vb,Holopainen:2011pd,Tuchin:2012mf,Linnyk:2013wma,Shen:2013vja,Basar:2012bp,Chatterjee:2013naa} though some open problems still remain. The theoretical calculations usually focus on the momentum spectrum of the radiation and thus treat photons as plane waves \cite{Kapusta:2006pm}. It has recently been pointed out in   
\cite{Chernodub:2010ye,Hirono:2015rla} that the spherical waves of photons, i.e.\ states of electromagnetic field with definite angular momentum, is an invaluable instrument for studying the topological properties of magnetic fields in media with a chiral anomaly, such as QGP. Certain photon spherical waves are states with definite magnetic helicity, which is a topological invariant  proportional to the number of twisted and linked flux tubes. These topological states are known as the Chandrasekhar-Kendall (CK) states \cite{CK,Biskamp}. 

In \cite{Hirono:2015rla} time-evolution of an initial topological state has been followed using the Chern-Simons-Maxwell model \cite{Wilczek:1987mv,Carroll:1989vb,Sikivie:1984yz,Kharzeev:2009fn}. 
The non-trivial physics of this evolution has been emphasized before in \cite{Tuchin:2014iua,Manuel:2015zpa,Joyce:1997uy,Akamatsu:2013pjd} in different contexts. The precise form of the initial state does not play an important role at long enough times. However,  the evolution time in QGP is restricted by its lifetime. This is why the initial condition must play an important role in the dynamics of electromagnetic fields.  

These observations motivate the author to address the problem of radiation of topological spherical  waves of photons, or simply, the CK photons. There are many ways to radiate a CK photon from the QGP. In this paper, which I consider as a benchmark for future studies, I discuss radiation of the CK photons by an ultrarelativistic quark scattering off thermal gluons: $q+g\to q+ \gamma_\text{CK}$. 
This process is similar to  Compton scattering except that the radiated photon is now in a topologically non-trivial state. The main goal of  this paper is to calculate the spectrum of the CK photons  emitted by an ultrarelativistic quark and study its properties.  

It is important to realize that the CK photon emission and the time-evolution of the magnetic field mentioned above do not interfere since they occur at different time scales. At long time scales, neither energy nor magnetic helicity are  conserved due to energy dissipation through the induced Ohmic electric currents. Moreover, magnetic helicity is non-conserved due to the anomalous (non-dissipative) electric current  generated by chiral imbalance in presence of the chiral anomaly \cite{Kharzeev:2009pj,Fukushima:2008xe}. However, these effects are small on the short time scales inherent for perturbative processes discussed in this paper.

The paper is structured as follows.  In  \sec{sec:b1} the spherical photon waves and the CK photons are introduced. This is followed by the quantization the electromagnetic field in the basis of the CK states in \sec{sec:b2} and  discussion of the applicability of the free-field approximation in \sec{sec:e}. Since calculation of the scattering matrix is most convenient in the momentum space, one needs to expand wave functions of the CK photons in the plane wave basis. This done in \sec{sec:c}. The main section is \sec{sec:d} where the scattering cross section is computed. Finally, in \sec{sec:f} the results are discussed and summarized.

\section{The CK states in the QGP}\label{sec:b1}

I am working in the radiation gauge $A_0=0$, $\b\nabla\cdot \b A=0$ in which $\b A$ satisfies the wave equation 
\ball{a21}
\nabla^2\b A -\partial_t^2\b A=0\,.
\gal
Its positive-energy solutions that have definite values of angular momentum can be written in the form
\ball{b19}
\b A_{klm}^h(\b r, t)= h k\b W^h_{klm}(\b r) e^{-i\omega_k t }\,,
\gal
where $\b W^{h}_{klm}(\b r)$ are eigenfunctions of the curl operator:
\ball{b23}
\b \nabla\times \b W^h_{klm}(\b r) = h k \b W^h_{klm}(\b r)\,.
\gal
Here $h= \pm 1$ is magnetic helicity, $l$ is the orbital angular momentum and $m=-l,\ldots,l$ its projection. The set of functions $\b W^{h}_{klm}(\b r)$ is complete on a unit sphere at any given $k$.
Their explicit expressions  read \cite{Jackson:1998nia}
\ball{b55}
\b W^h_{klm}(\b r) = \bar N_k\left(\b T^h_{klm}(\b r)-ih \b P^h_{klm}(\b r)\right)\,,
\gal
where 
\ball{b57}
\b T^h_{klm}(\b r)= \frac{j_l(kr)}{\sqrt{l(l+1)}}\b L[Y_{lm}(\theta,\phi)]\,,\quad \b P^h_{klm}(\b r)= \frac{i}{k}\b \nabla \times \b T^h_{klm}(\b r)\,, \quad l\ge 1\,.
\gal
Here  $\b L = -i(\b r\times \b \nabla)$ is the orbital angular momentum operator. Although functions $\b T_{klm}$ and $\b P_{klm}$ also form a complete set on a unit sphere (at fixed $k$), they do not  have definite magnetic helicity. The normalization constant $\bar N_k$ is  given by
\ball{b61}
\bar N_k =\frac{1}{\sqrt{2}\,\omega_k^{3/2}}\cdot \left\{\begin{array}{cl}\frac{1}{R^{3/2}j_{l+1}(kR)}\,, & \text{discrete}\,; \\ k\,, & \text{continuous}\,,\end{array}\right.
\gal
for  the discrete and continuous spectra respectively. It is chosen in such a way that the orthogonality conditions read
\ball{b24}
\int \b W^{h'*}_{k'l'm'}(\b r)\cdot \b W^{h}_{klm}(\b r)d^3 r= \frac{1}{2\omega_k^3}\delta_{kk'}\delta_{ll'}\delta_{mm'}\delta_{h h'}\,,
\gal
for a discrete spectrum  and 
\ball{b24a}
\int \b W^{h'*}_{k'l'm'}(\b r)\cdot \b W^{h}_{klm}(\b r)d^3 r=  \frac{\pi}{2 \omega_k^3}\delta(k-k')\delta_{ll'}\delta_{mm'}\delta_{h h'}\,,
\gal
for a continuous one, which can be readily verified using the properties of the spherical Bessel functions 
\bal
&\int_0^R j_l(kr)j_l(k'r)r^2 dr = \frac{1}{2}R^3\delta_{kk'}j_{l+1}^2(kR)\,, \label{b65a}\\
&\int_0^\infty j_l(kr)j_l(k'r)r^2 dr =\frac{\pi}{2k^2}\delta(k-k')\,.\label{b65b}
\gal
 The normalization conditions \eq{b24},\eq{b24a} are used in the relativistic scattering theory (though they are different from \cite{Hirono:2015rla} and \cite{Jackson:1998nia}).

A discrete spectrum with the quantized  $k$ emerges if one imposes a boundary condition on the magnetic field at some distance $r=R$ \cite{Chernodub:2010ye}. Although there seems to be no physical reason to impose such a boundary condition, in practical calculations it is sometimes more convenient to handle a discrete spectrum than  a continuous one, and afterwards set $R\to \infty$. Throughout the paper the same letter $k$ is used to denote both the continuous and discrete variables.

\section{Quantization of the free field}\label{sec:b2}

Substituting \eq{b19} into \eq{a21} we obtain the dispersion relation of free photons: $\omega_k=k$.
In terms of the normal modes \eq{b19}, the vector potential reads (in the discrete case)
\ball{b30}
\b A(\b r, t)= \sum_{klmh}\left( h k a_{klm}^h\b W_{klm}^h(\b r) e^{-i\omega_k t}+c.c.\right)\,,
\gal
where $a_{klm}^h$ are arbitrary coefficients. The electric and magnetic fields are given by
\bal
\b E(\b r, t)&=-\partial_t \b A(\b r,t)= \sum_{klmh}\left( ih k\omega_k a_{klm}^h\b W_{klm}^h(\b r) e^{-i\omega_k t} +c.c.\right)\,,\label{b35}\\
\b B(\b r, t)&= \b \nabla\times \b A(\b r, t)=\sum_{klmh}\left(  k^2 a_{klm}^h\b W_{klm}^h(\b r) e^{-i\omega_k t} +c.c.\right)\label{b36}\,,
\gal
The total electromagnetic energy of the discrete spectrum can be written as a sum over the energies of all CK states as follows: 
\ball{b37}
\mathcal{E}= \frac{1}{2}\int (\b E^2+\b B^2)d^3r= \sum_{klm}\omega_k\, a_{klm}^h a_{klm}^{h*}\,.
\gal
The normalization in \eq{b24} and \eq{b24a} was chosen so that the energy of a single CK state is $\omega_k$.   For the continuous spectrum, the total electromagnetic energy  can be written as\footnote{Coefficients $a_{klm}^h$ are normalized differently in \eq{b37} and \eq{b46}.}
\ball{b46}
\mathcal{E}=\sum_{lmh}\int _0^\infty \frac{dk}{\pi} |\omega_k| a_{klm}^h a_{klm}^{h*}\,.
\gal
The magnetic helicity of electromagnetic field reads
\ball{b48}
H= \int \b A\cdot \b B\, d^3r=  \sum_{klmh}h\,a_{klm}^h a_{klm}^{h*}\,,
\gal
indicating that magnetic helicity of a single CK state is $h$. Clearly, it is a conserved quantity.

Finally, the quantized electromagnetic field can be written down using \eq{b30} as 
\ball{b73}
\b A(\b r, t)= \sum_{klmh}\left(h  k a_{klm}^h\b W_{klm}^h(\b r) e^{-i\omega_k t}+ h.c.\ \right)\,,
\gal
where now $a_{klm}^h$ is an operator obeying the usual bosonic commutation relations.

\section{Role of electric currents}\label{sec:e}

It is important to delineate the region of applicability of the free-field approximation of the previous section.  The perturbation theory that I am employing in this paper, hinges on the assumption that the electrical currents in the medium can be treated as small perturbations, since they  are proportional to $\alpha_\text{em}$.  For illustration, consider a model  of classical electrodynamics with an anomalous current given by the Maxwell-Chern-Simons equations \cite{Wilczek:1987mv,Carroll:1989vb,Sikivie:1984yz,Kharzeev:2009fn}:
\bal
&\b \nabla\cdot \b B=0\,, \label{b11}\\
& \b \nabla\cdot \b E= 0\,,  \label{b12}\\
& \b \nabla \times \b E= -\partial_t \b B\,,\label{b13}\\
& \b \nabla \times \b B= \partial_t \b E+ \sigma \b E + \sigma_\chi \b B\,.\label{b14}
\gal
In the last equation $\sigma \b E$ and $\sigma_\chi \b B$ stand for the Ohmic and anomalous electrical current densities.  In the radiation gauge  Eqs.~\eq{b11}-\eq{b14} yield an equation for the vector potential
\ball{b17}
-\nabla^2\b A= -\partial_t^2\b A-\sigma \partial_t \b A+\sigma_\chi\b\nabla\times \b A\,,
\gal
where for simplicity I treat $\sigma_\chi$ as a positive constant.
Substituting \eq{b19} into \eq{b17} one finds a dispersion relation 
\ball{b25}
k^2-h \sigma_\chi k -\omega_k(\omega_k+i\sigma)=0\,,
\gal
which has the following two solutions: 
\ball{b25a}
\omega_k= -\frac{i\sigma}{2}\pm \sqrt{(k^2-h\sigma_\chi k)-\sigma^2/4}\,.
\gal
Evidently, if $k\gg \sigma$, the dissipation effects due to the Ohmic currents can be neglected. 

Unlike the dissipative currents that preclude the very notion of definite energy states, the anomalous currents are non-dissipative \cite{Kharzeev:2013ffa} and in principle the electromagnetic field could have been quantized in their presence. However, there a two problems. First, the magnetic helicity is not conserved due to the chiral anomaly, see e.g.\ \cite{Hirono:2015rla}. Second, the dispersion relation \eq{b25a} contains an unstable solution at $k<\sigma_\chi$ \cite{Tuchin:2014iua,Manuel:2015zpa,Joyce:1997uy,Akamatsu:2013pjd}. This is easily seen at $\sigma=0$: when $h=1$, $\omega_k$ is purely imaginary and positive, hence the corresponding eigenfunction exponentially increases with time. Taking the time-dependence of $\sigma_\chi$ into account does not resolve the problem \cite{Joyce:1997uy}. In fact, the presence of a spatially uniform current $\sigma_\chi \b B$ in \eq{b14} violates causality. It seems possible that a more realistic model of the anomalous term may cancel the instability. These problems, however, are rather academic. Indeed, as far as the QGP of a realistic size $R\sim5-10$~fm and electrical conductivity $\sigma\sim 5-6$~MeV \cite{Ding:2010ga,Aarts:2007wj,Aarts:2014nba,Cassing:2013iz} is concerned, the requirement that $k\gg \sigma$ is always satisfied  because $k\gtrsim 1/R\gg \sigma$. Since $\sigma_\chi$ is probably of the same order of magnitude as $\sigma$, it implies that $k\gg \sigma_\chi$ and the unstable modes do not contribute to the spectrum of  the CK  states (and neither does the anomaly).

\section{Plane wave expansion of  the CK states }\label{sec:c}

The wave function of a photon  with a given momentum $\b q$ and polarization $\lambda$ is given by
\ball{c9}
\bm {\mathcal A}_{\b q\lambda}(\b r)= \frac{1}{\sqrt{2\omega V}}\b \epsilon_{\b q\lambda}e^{i\b q\cdot\b r}\,,
\gal
where $\lambda=1,2$ are two polarization states and $V$ is the plasma volume. In the next section I will need the following expansion of the CK states into the plane waves \eq{c9}:
\ball{c11}
\b W_{klm}^h(\b r)= \sum_\lambda \int \frac{d^3q}{(2\pi)^3} e^{i\b q\cdot\b r}\b \epsilon_{\b q\lambda} w_{klm}^h(\b q,\lambda)\,.
\gal
where
\ball{c13}
w_{klm}^h(\b q,\lambda)= \int d^3r\, e^{-i\b q\cdot \b r}\b\epsilon^*_{\b q\lambda}\cdot  \b W^h_{klm}(\b r)=\bar N_k\b\epsilon^*_{\b q\lambda}\cdot\left(\b T^h_{klm}(\b q)-ih \b P^h_{klm}(\b q) \right)
\gal
with
\ball{c15}
\b T^h_{klm}(\b q) = \int d^3r\, e^{-i\b q\cdot \b r}\b T^h_{klm}(\b r)\,,\quad 
\b P^h_{klm}(\b q) = \int d^3r\, e^{-i\b q\cdot \b r}\b P^h_{klm}(\b r)\,.
\gal
Substituting \eq{b57} and denoting by $\unit r$ a unit vector in the $\b r$ direction yields
\bal
\b T^h_{klm}(\b q) &= \int d^3r\, e^{-i\b q\cdot \b r}\frac{j_l(kr)}{\sqrt{l(l+1)}}\b L[Y_{lm}(\unit r)] = -\int d^3r\, \b L[e^{-i\b q\cdot \b r}]\frac{j_l(kr)}{\sqrt{l(l+1)}}Y_{lm}(\unit r) \label{c19}\\
&= \b L \int d^3r\, e^{-i\b q\cdot \b r}\frac{j_l(kr)}{\sqrt{l(l+1)}}Y_{lm}(\unit r) \,,
\label{c20}
\gal
where in the last line the angular momentum operator is in the momentum representation: $\b L = -i\b q\times \b \nabla_{\b q}$. Using the expansion of the plane wave into the spherical waves
\ball{c23}
e^{i\b q\cdot \b r}= 4\pi \sum_{lm}i^l j_l(qr)Y_{lm}^*(\unit q)Y_{lm}(\unit r)\,,
\gal
along with the orthonormality relations \eq{b65a},\eq{b65b} furnishes 
\ball{c25}
\b T^h_{klm}(\b q) &=2\pi R^3(-i)^l \frac{j_{l+1}^2(kr)}{\sqrt{l(l+1)}}\delta_{kq}\, \b L Y_{lm}(\unit q)\,,
\gal
for the discrete spectrum and 
\ball{c26}
\b T^h_{klm}(\b q) &=\frac{2\pi^2}{ k^2}(-i)^l \frac{1}{\sqrt{l(l+1)}} \delta(k-q)\,\b L Y_{lm}(\unit q)
\gal
for the continuous one. From \eq{b57} and \eq{c15} it follows that
\ball{c29}
\b P^h_{klm}(\b q)= -\frac{1}{k}\b q\times \b T^h_{klm}(\b q)\,.
\gal
Substituting \eq{c25}--\eq{c29} into \eq{c13} we obtain the desired expansion of the CK states into the plane waves.

\section{Spectrum of the CK photons}\label{sec:d}

Consider an ultra-relativistic quark traveling through  the QGP at temperature $T$ with energy $\e\gg T$. As it scatters off a thermal gluon it radiates a CK photon through the process  $q(p^\mu)+g(k^\mu)\to q(p'^\mu)+\gamma_\text{CK}(k'lmh)$. Quantities in parentheses denote the quantum numbers of the corresponding particles;  in the case of  quarks and gluons these are their 4-momenta. The scattering matrix element for this process reads
\ball{d3}
S_{fi}=&e_qg \int d^4x\int d^4 y \,\bar \psi_f(x)\left[  i\slashed{A}_{k'lm}^{h*}(x)\, iS(x-y) (-i \slashed{A}^a_{k\lambda}(y))\right. \nonumber\\
 &  \left. + (-i\slashed{A}^a_{k\lambda}(x))\, iS(x-y) (i \slashed{A}_{k'lm}^{h*}(y))  \right]\psi_i(y)\,,
\gal
where $e_q$ and $g$ are the electromagnetic and strong coupling constants (assumed to be small).  The gluon field potential $A^a_{k\lambda}(y)$ is normalized as in \eq{c9}:
\ball{d5}
A^a_{ k\lambda}(\b r)= \frac{1}{\sqrt{2\omega V}}t^a\epsilon_{k\lambda}e^{-i k_\mu\cdot x^\mu}\,,
\gal
where $t^a$ is the color symmetry generator. The initial and final quark wave functions are states with given momentum and spin that read
\bal
\psi_i(y)&= \frac{1}{\sqrt{2\e V}}\, u_{ps}\, e^{-i p_\mu\cdot y^\mu}\label{d7}\\
\bar\psi_f(x)&= \frac{1}{\sqrt{2\e' V}}\, \bar u _{p's'}\, e^{ip_\mu'\cdot x^\mu}\label{d8}\,,
\gal
where $s$ and $s'$ stand for spin projections. The most convenient way to compute the scattering matrix element \eq{d3} is to employ the plane wave representation of  the CK photon wave function derived in the  previous section.  Substituting  \eq{c11} into  \eq{b19} we have
\ball{d11}
\b A_{k'lm}^{h}(x)= h k'\b W^h_{k'lm}(\b x) e^{-i\omega_k' x_0 }= h k'\sum_{\lambda'} \int \frac{d^3q'}{(2\pi)^3} e^{-iq'_\mu\cdot x^\mu}\b \epsilon_{\b q'\lambda'} w_{k'lm}^h(\b q',\lambda')\,,
\gal
where I denoted $q'^\mu=(\omega_q',\b q')$. Energy conservation, which is explicit in  \eq{c25},\eq{c26}, requires $q'=k'$ implying that $\omega_q'= \omega_k'$. Substituting \eq{d5},\eq{d7},\eq{d8},\eq{d11}  into \eq{d3} and using the momentum space representation of the free fermion propagator 
\ball{d17}
S(x-y)= \int\frac{d^4\ell}{(2\pi)^4}e^{-i\ell \cdot (x-y)}\frac{\slashed{\ell}+m_q}{\ell^2-m_q^2+i\epsilon}\,,
\gal
one arrives at the following expression:
\ball{d19}
S_{fi}= \frac{-i (2\pi) h k't^a}{\sqrt{2\e'}\sqrt{2\e}\sqrt{2\omega}V^{3/2}}\sum_{\lambda'}\int d^3 q'\, w_{k'lm}^{h*}(\b q',\lambda')\,\mathcal{M}_C(qg\to q\gamma)\,\delta^4(p^\mu+k^\mu-p'^\mu-q'^\mu)\,,
\gal
where  the Compton scattering amplitude is given by:
\ball{d21}
\mathcal{M}_C= -e_q\,g\,\bar u_{p's'}\left[\slashed{\epsilon}_{q\lambda'}^* \frac{\slashed{k}+\slashed{p}+m_q}{(k+p)^2-m_q^2+i\epsilon}\slashed{\epsilon}_{k\lambda} +
\slashed{\epsilon}_{k\lambda}\frac{\slashed{p}-\slashed{q}'+m_q}{(p-q')^2-m_q^2+i\epsilon}
\slashed{\epsilon}_{q\lambda'}^*\right]u_{ps}\,
\gal
and describes the Compton scattering $q(p^\mu)+g(k^\mu)\to q(p'^\mu)+\gamma(q'^\mu)$ (modulo the color matrix $t^a$ that is extracted for convenience). The scattering matrix in \eq{d19} is a convolution of  two physical processes: the Compton scattering yielding an intermediate photon carrying momentum $q'^\mu$ and polarization $\lambda'$ and the conversion of this photon into a CK state.  Integrating over $\b q'$ in \eq{d19} and multiplying by its complex conjugate  one finds 
\ball{d24}
|S_{fi}|^2= \frac{(2\pi)^2 k'^2 t^at^a}{2\e'\,2\e\,2\omega\,V^3}\delta(\e+\omega-\e'-\omega_k')\frac{t}{2\pi}\left|\sum_{\lambda'} w_{k'lm}^{h*}(\b p+\b k-\b p',\lambda')\,\mathcal{M}_C(qg\to q\gamma)\right|^2\,,
\gal
where $t$ is the total observation time. The production rate of a CK photon is calculated as follows:
\ball{d29}
\mathcal{R}_{lm}^h(k')= \frac{1}{4}\frac{1}{2N_c} \frac{(2\pi) k'^2}{2\e'\,2\e\,2\omega\, V}\int \sum_{\lambda s s'}\left|\sum_{\lambda'} w_{k'lm}^{h*}(\b p+\b k-\b p',\lambda')\,\mathcal{M}_C(qg\to q\gamma)\right|^2&\nonumber\\
\times\delta(\e+\omega-\e'-\omega_k')\frac{d^3p'}{(2\pi)^3}f(\omega)\frac{d^3k}{(2\pi)^3}\,,&
\gal
where the Compton color factor is $\frac{1}{N_c(N_c^2-1)}\Tr(t^at^a)= \frac{1}{2N_c}$. The  Bose-Einstein distribution of gluons  in the QGP at temperature $T$ is given by
\ball{d31}
f(\omega) = \frac{2(N_c^2-1)}{e^{\omega/T}-1}\,.
\gal
Eq.~\eq{d29} gives  the CK photon production rate in the case of a discrete spectrum.  To obtain a formula for the continuous spectrum, the left-hand-side of \eq{d29} should be replaced by the rate density $\mathcal{R}\to \pi d\mathcal{R}/dk'$.  Since $|\b k|= \omega$ one can use the remaining delta-function to take the integral over the gluon momentum to derive (for a continuous spectrum)
\ball{d33}
\pi\frac{d\mathcal{R}_{lm}^h}{dk'}= \frac{1}{4}\frac{1}{2N_c} \frac{(2\pi) k'^2}{2\e'\,2\e\,2\omega\, V}\sum_{\lambda s s'}\left|\sum_{\lambda'} w_{k'lm}^{h*}(\b q',\lambda')\,\mathcal{M}_C(qg\to q\gamma)\right|^2\frac{d^3q'}{(2\pi)^3}f(\omega)\frac{d\Omega_{\b k} \omega^2}{(2\pi)^3}\,,
\gal
where $d\Omega_{\b k} $ is an element of the solid angle in the gluon momentum $\b k$ direction.
I also changed the integration variable form $\b p'$ to $\b q' = \b p+\b k-\b p'$.
To obtain the cross section one has to divide the rate by the flux density
\ball{d34}
j = \frac{p\cdot k }{V \e \omega}\,.
\gal

In the ultrarelativistic limit  the quark mass $m_q$ can be neglected. In this case  the Compton amplitude $\mathcal{M}_C$ is  the same for the  left- and right-polarized gluons (see e.g.\ \cite{Peskin:1995ev}). Thus, it can be taken out of the sum over the intermediate photon polarizations  $\lambda'$ in \eq{d29}:
\ball{d39}
\left|\sum_{\lambda'} w_{k'lm}^{h*}(\b q',\lambda')\,\mathcal{M}_C\right|^2\approx |\bar N_{k'}|^2\left| \mathcal{M}_C\right|^2\left|\sum_{\lambda'}\b\epsilon_{\b q'\lambda'}\cdot\left(\b T^{h*}_{k'lm}(\b q')+ih \b P^{h*}_{k'lm}(\b q') \right)\right|^2\,,
\gal
where  \eq{c13} was used. The Compton amplitude is obtained with the standard manipulations \cite{Peskin:1995ev}:
\ball{d41}
\sum_{\lambda s s'}\left| \mathcal{M}_C\right|^2=\frac{1}{2}\sum_{\lambda\lambda' s s'}\left| \mathcal{M}_C\right|^2= 4e_q^2g^2\,\left( \frac{p\cdot k}{p\cdot q'}+ \frac{p\cdot q'}{p\cdot k} \right)= 8e_q^2g^2\,,
\gal
where in the last equation I used the conservation law $(p'^\mu)^2 = (p^\mu+k^\mu-q'^\mu)^2$ that implies that $p\cdot k = p\cdot q'$ for a highly energetic quark $\e\gg \omega$ \footnote{In principle, if the scattering angle $\chi$ is very small $\chi\ll m_q/\e$, the scattered quark energy $\e'$ can be comparable to the energy of  a CK state $\omega_k'$. However, as we will demonstrate below, the main contribution to the cross section comes from $\omega_k'< T\ll \e$.}. Let $\chi$ be the angle between the vectors $\b p$ and $\b q'$ and $\theta$  be the angle between the vectors $\b p$ and $\b k$. Then $  p\cdot q'=p\cdot k$ implies 
\bal
\omega_q'(1-\cos\chi)= \omega (1-\cos\theta)\label{d43}\,.
\gal
 Substituting \eq{d39} and \eq{d41} into \eq{d33} and dividing by the flux density \eq{d34} one gets the differential cross section:
\ball{d47}
\pi\frac{d\sigma_{lm}^h}{dk'}=& \frac{e_q^2g^2}{8(2\pi)^5}\frac{ k'^2\omega}{N_c \e'\e(1-\cos\theta)}|\bar N_{k'}|^2f(\omega) \left|\sum_{\lambda'}\b\epsilon_{\b q'\lambda'}\cdot\left(\b T^{h*}_{klm}(\b q')+ih \b P^{h*}_{klm}(\b q') \right)\right|^2  d^3q' d\Omega_{\b k}\,.
\gal
 To integrate over the directions of $\b k$ it is helpful to  introduce a new variable $y= 1/(1-\cos\theta)$  and denote $a=  \omega_q'(1-\cos\chi)$. In view of \eq{d43} one finds 
\ball{d49}
\int \frac{\omega\, d\Omega_{\b k}}{(e^{\omega/T}-1)(1-\cos\theta)} =
2\pi a\int_{1/2}^\infty \frac{dy}{e^{a y /T}-1}
=
-2\pi T\ln (1-e^{-a/2T})\,.
\gal
To sum over the polarizations $\lambda'$ of the intermediate photon introduce a Cartesian coordinate system $\xi\eta\zeta$ with $\zeta$-axis in the direction of  vector $\b q'$. Since the factorization property \eq{d39} holds only for circularly  polarized photons we choose the polarization vectors as $\b\epsilon_{\b q'\lambda'}= \frac{1}{\sqrt{2}}(\unit \xi+i\lambda'\unit \eta)$, with $\lambda'=\pm 1$. This yields (for a continuous spectrum):
\bal
&\left|\sum_{\lambda'}\b\epsilon_{\b q'\lambda'}\cdot\left(\b T^{h*}_{k'lm}( \b q')+ih \b P^{h*}_{k'lm}(\b q') \right)\right|^2= 2\left|  \unit \xi \cdot \left(\b T^{h*}_{k'lm}(\b q')+ih \b P^{h*}_{k'lm}(\b q') \right)\right|^2\label{d55}\\
=&\frac{8\pi^4}{k'^4 l(l+1)}[\delta(k'-q')]^2\left|  \left(L_\xi +ihL_\eta\right) Y^*_{lm}(\unit q')  \right|^2\,, \label{d56}
\gal
where I used \eq{c26} and \eq{c29}. To integrate over $\b q'$, introduce another Cartesian coordinate system $xyz$, such that $z$-axis is in the direction of vector $\b p$. The direction of $\b q'$ in this frame is characterized by the polar and azimuthal angles $\chi$ and $\phi$. Without loss of generality fix $\unit\eta$ to be in the plane of vectors $\b p$ and $\b q'$. The two coordinate frames are related as follows:
\bal
\unit \xi & = \sin\phi\,\unit x- \cos\phi\, \unit y\,,\label{d60a}\\
\unit \eta &= \cos\chi \cos\phi\, \unit x+\cos \chi\sin\phi\,\unit y -\sin\chi \,\unit z\,,\label{d60b}\\
\unit \zeta &= \sin\chi \cos\phi \,\unit x+\sin \chi\sin\phi \,\unit y +\cos\chi\, \unit z\,.\label{d60c}
\gal
Using these formulas it is straightforward to derive
\ball{d62}
-i(L_\xi +ihL_\eta)= \frac{1}{2}e^{-i\phi}(1+h\cos\chi)L_++\frac{1}{2}e^{i\phi}(-1+h\cos\chi)L_--h\sin\chi L_z\,,
\gal
where 
\ball{d64}
L_\pm Y_{lm}&= (L_x\pm  iL_y)Y_{lm}= \sqrt{l(l+1)-m(m\pm 1)}\,Y_{l,m\pm 1}\,.
\gal
To write the cross section in a compact form define the $F$-function
\ball{d67}
F_{lm}^h(x)=- \int  \ln (1-e^{-x(1-\cos\chi)/2})\frac{1}{l(l+1)}\left|  \left(L_\xi -ihL_\eta\right) Y_{lm}(\unit q')  \right|^2    d\Omega_{\b q'}\,.
\gal
Since $\phi$-dependence of the spherical harmonics is given by $Y_{lm}\sim e^{im\phi}$,  it follows from \eq{d62} and \eq{d64}, that function $(L_\xi -ihL_\eta)Y_{lm}$ contains three terms that depend on $\phi$ as $e^{i(m+2)\phi}$, $e^{-i(m+2)\phi}$  and $e^{i\phi}$. These are mutually orthogonal on a unit circle meaning that the three terms do not interfere once we integrate over the directions of $\b q'$ in \eq{d47}. This brings the $F$-function to the following form:
\ball{d72}
&F_{lm}^h(x)= -\frac{2\pi}{l(l+1)}\int_0^\pi d\chi \sin\chi \ln (1-e^{-x(1-\cos\chi)/2})\nonumber\\
&\times 
\left\{  \sin^2\chi \, m^2|Y_{lm}(\chi,0)|^2      +\frac{1}{4}(1-h\cos\chi)^2\left[l(l+1)-m(m-1)\right]|Y_{l,m-1}(\chi,0)|^2\right.\nonumber\\
&\left.  +\frac{1}{4}(1+h\cos\chi)^2\left[l(l+1)-m(m+1)\right] |Y_{l,m+1}(\chi,0)|^2 \right\}   \,.
\gal
Explicit expressions for the the $p$-wave  $F$-functions are listed in the Appendix. 
Substituting \eq{d56} and \eq{d49} into \eq{d47} and replacing one of the delta functions by $R/\pi$ one obtains for the continuous spectrum
\ball{d73}
\frac{1}{R}\frac{d\sigma_{lm}^h}{dk'} = \frac{e_q^2g^2C_F}{16\pi^2}\frac{T}{k'\e^2}F_{lm}^h(k'/T)\,,
\gal
which is the radiation cross section per unit length, while for the discrete spectrum 
\ball{d74}
\sigma_{lm}^h= \frac{e_q^2g^2}{16\pi}\frac{C_F}{R^2 j^2_{l+1}(k'R)}\frac{ T}{k'^3\e^2}F_{lm}^h(k'/T)\,,
\gal
where $C_F= (N_c^2-1)/2N_c$ is a color Casimir invariant.

\begin{figure}[ht]
\begin{tabular}{cc}
      \includegraphics[height=5cm]{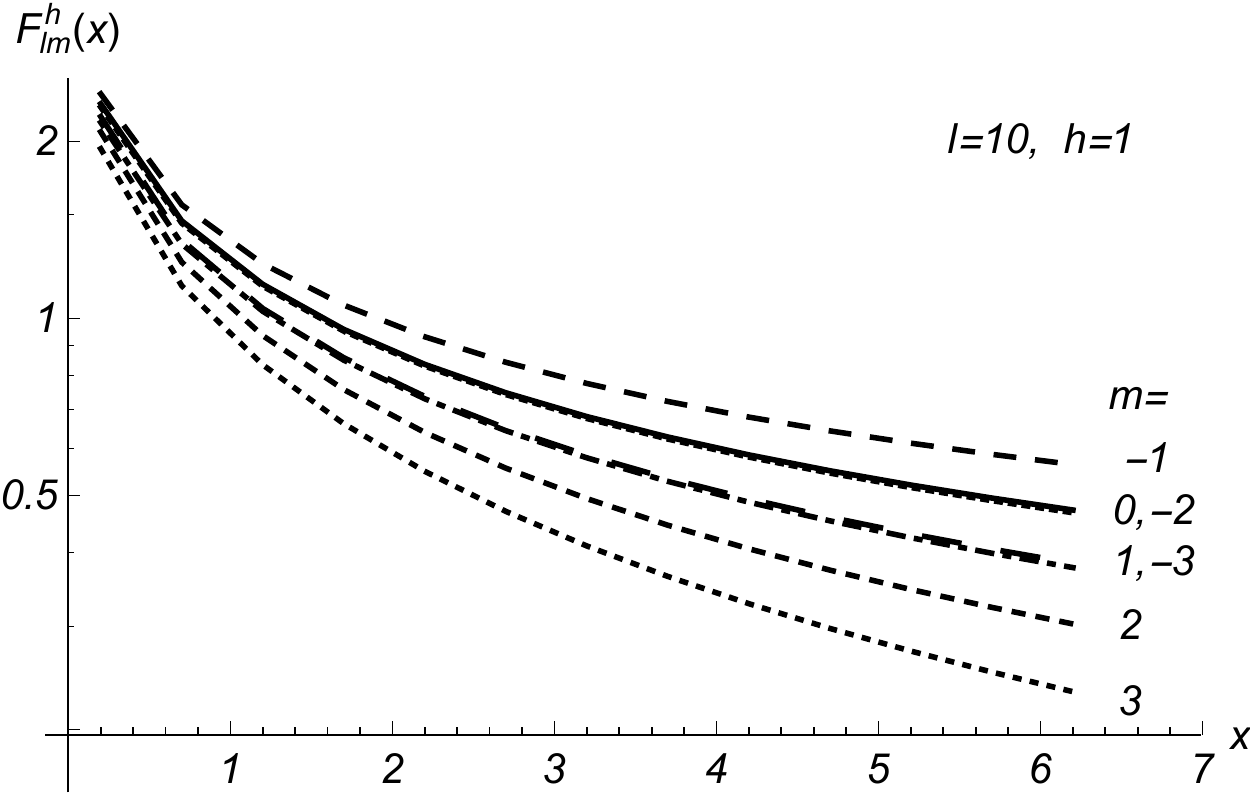} &
      \includegraphics[height=5cm]{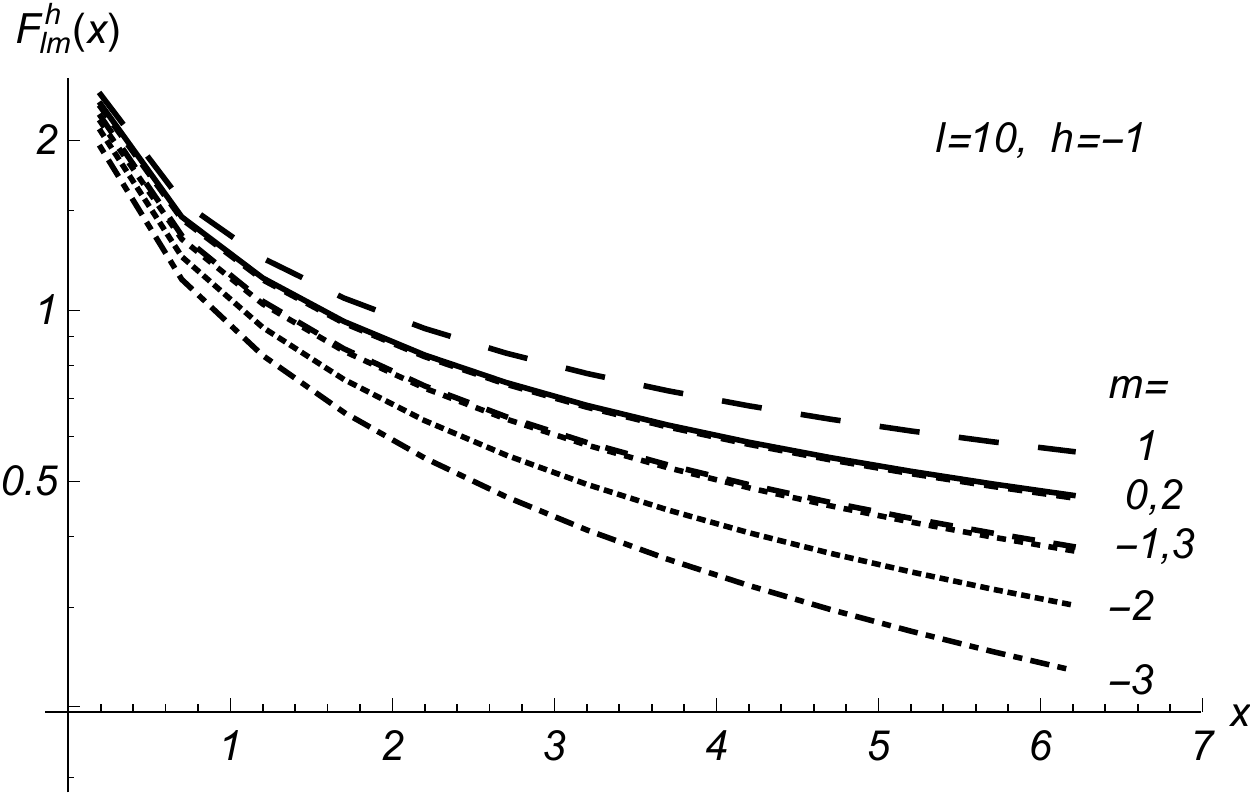}\\
      $(a)$ & $(b)$ 
      \end{tabular}
  \caption{Function $F^h_{lm}(x)$ at $l=10$ and different $m$. Left panel: $h=1$, right panel: $h=-1$. Variable $x$ is the CK photon energy in units of the plasma temperature. }
\label{fig1}
\end{figure}

In the long wavelength limit  $k'\ll T$, which corresponds to small values of $x$, the logarithm under the integral in \eq{d72} is approximately $\chi$-independent implying that $F$ is proportional to $\ln(1/x)$, with the proportionality coefficient being dependent on $h$, $l$ and $m$. In the other limit $x\gg 1$ the logarithm under the integral in \eq{d72} is approximated by $e^{-x(1-\cos\chi)/2}$ implying that the integrand is exponentially suppressed unless $\chi$ is very small $\chi\lesssim 2/\sqrt{x}$. Thus, at large $x$ the main contribution comes from $\chi\ll 1$. Expansion of the spherical harmonics to the order $\mathcal{O}(\chi^2)$ reads:
\ball{d80}
Y_{lm}(\chi,\phi)=& \sqrt{\frac{2l+1}{4\pi}}\left[\delta_{m0}\mp \frac{e^{\pm i\phi}}{2}\sqrt{l(l+1)} \delta_{m,\pm 1}\chi  -\frac{1}{4}l(l+1)\delta_{m0}\chi^2\right.\nonumber\\
&\left.\mp \frac{e^{\pm 2i\phi}}{8}\sqrt{(l+2)(l+1)l(l-1)}\delta_{m,\pm 2}\chi^2
\right]\,.
\gal
Expanding the integrand of \eq{d72} to the leading order in $\chi$ yields
\ball{d82}
F_{lm}^h(x)\approx& -\frac{2\pi}{l(l+1)}\int_0^{\infty} d\chi\, \chi \ln \left( 1-e^{-x\chi^2/4}\right)[l(l+1)-m(m+h)]\frac{2l+1}{4\pi}\delta_{m,-h}\nonumber\\
=& \frac{\pi^2}{6} \frac{1}{x}(2l+1)\delta_{m,-h}\,, \quad x\gg 1\,.
\gal
One concludes that the largest partial wave of a CK photon at high energies $k'$ has orbital angular momentum projection $m=\pm 1$ and magnetic helicity $h=\mp 1$. It is not difficult to verify by expanding spherical harmonics at small $\chi$ in \eq{d82} and keeping the higher order terms, that the function $F$ for a partial wave with given $m$ and $h$ depends on $x$ as $1/x^{|m+h|+1}$ at $x\gg 1$. This can be seen on \fig{fig1} and in \eq{app7}. It is evident  from \fig{fig1} that most of the CK photons have wavelengths larger than $2\pi/T$.  

The total cross section diverges at large $l$ as $l^2$. The maximum possible value of $l$ can be determined from the condition $k'\ll \e$. In the discrete spectrum case $k'R= x_{l,n}$, where $x_{l,n}$ is  the $n$'th zero of the function $j_l(x)$. According to 9.5.22 in \cite{AS} at large $l$, $x_{l,n}\approx l$ implying that $l\ll \e R$. Since the spectrum is largest at $x<1$, the typical values of the angular momenta are $l< TR\sim 5-10$ for a realistic system. The total cross section is independent of the quark's energy and is proportional to the plasma volume and temperature.

\section{Summary and outlook }\label{sec:f}

The main result of this paper is Eqs.~\eq{d72}--\eq{d73}, which represent the spectrum of  the CK photons radiated by an ultra-relativistic quark $\e\gg T$ moving through the QGP. The CK photon spectrum decreases rather slowly with photon energy $k'$ as $k'^{-1}\ln (T/k')$ at low energies $k'\ll T$ and as  $1/k'^{|m+h|+2}$ at high energies $k'\gg T$. The average energy of a CK photon is $\aver{k'}= 2T/\ln^2(RT)$,  while the average wavelength is $\pi/\aver{k'}$. Numerically it is of the order of the QGP linear size $R$, where I used $T=200$~MeV and $R=5$~fm. The production rate is highest for states with $m=1$, $h=-1$ and $m=-1$, $h=1$.  The CK photon spectrum  falls off as $1/\e^2$ as the quark energy increases, indicating that the contribution of the thermal quarks, which carry energies $\e\sim T$,  maybe essential. This entails taking recoil effect into account, which is conceptually straightforward, but makes  equations bulky. 

 In a medium with the chiral imbalance of right and left-handed quarks $N_R$ and $N_L$ the spectrum of the CK photons will be asymmetric  with respect to $m\to -m$ transformation. In particular, if $N_R>N_L$ the $m=-1$ component of the spectrum will be enhanced as compared to $m=1$ which can serve as a signature of the  chiral imbalance.

There are a number of assumptions concerning  the QGP that I made in order to reach the  analytical result \eq{d72}--\eq{d74}.  I considered a model in which the QGP is stationary and  spatially uniform, which allowed me to handle the problem analytically. However, even in a stationary medium,  chiral conductivity $\sigma_\chi$ has important  temporal and, possibly, spatial dependence. In particular, it has been recently argued that disturbing  the QGP out of the chiral equilibrium, initiates a relaxation process in which electromagnetic field and fermions exchange helicity \cite{Hirono:2015rla,Manuel:2015zpa,Joyce:1997uy}. Nevertheless, this process takes place on time scales much larger than the typical scattering process in plasma.  It has also been tacitly assumed that  the QGP temperature be high enough to allow application of  the perturbation theory.  

Eqs.~\eq{d72}--\eq{d73} provide an initial condition for the  topological evolution of the electromagnetic field  discussed recently in \cite{Hirono:2015rla}. In contrast to \cite{Hirono:2015rla}, which assumed an exponential spectrum, it is found that the photon spectrum is given by a power law.  The photon spectrum computed in this paper may also be modified as the photons propagate though the QGP in the presence of the intense external electromagnetic field \cite{Tuchin:2013ie}. This and other related issues will be addressed  elsewhere.

\acknowledgments
The author wishes to  thank Dima Kharzeev  for helpful communications.
This work  was supported in part by the U.S.\ Department of Energy under Grant No.\ DE-FG02-87ER40371.

\appendix
\section{Explicit expressions of the $p$-wave  $F$-functions.}\label{appA}

The $F$-functions defined in  \eq{d72} can be expressed in terms of the polylogarithm functions $\text{Li}_s(e^{-x})$. For example, at $l=1$ they read
\bal
F_{10}^{\pm1}(x)=&\frac{6}{x^5}\left[ 48 \text{Li}_6(e^{-x})+24x \text{Li}_5(e^{-x})+ 6x^2 \text{Li}_4(e^{-x})+x^3\text{Li}_3(e^{-x})+x^3\zeta(3)+24x \zeta(5) \right]\nonumber\\
&-\frac{2\pi^4}{105 x^5}(16\pi^2+21 x^2)\,, \label{app1}\\
F_{1, \pm 1}^{\pm 1}(x)=&-\frac{3}{x^5}\left[ 192 \text{Li}_6(e^{-x}) + 120 x\text{Li}_5(e^{-x})+34x^2 \text{Li}_4(e^{-x})+6x^3 \text{Li}_3(e^{-x})+x^4 \text{Li}_2(e^{-x})\right. \nonumber\\
&\left. +72x\zeta(5)\right] +\frac{\pi^4(64\pi^2+35x^2)}{105 x^5}\,,  \label{app2}\\
F_{1, \pm 1}^{\mp 1}(x)=&-\frac{6}{x^5}\left[ 96 \text{Li}_6(e^{-x}) + 36 x\text{Li}_5(e^{-x})+5x^2 \text{Li}_4(e^{-x})+3x^3\zeta(3)+60 x \zeta(5)\right]\nonumber\\
&+\frac{\pi^2(128\pi^4+238\pi^2 x^2+ 105 x^4)}{210 x^5}\,, \label{app3}
\gal
where $\zeta(n)$ is the Riemann zeta function. Asymptotic expressions at low $x$ are:
\ball{app5}
F_{10}^{\pm1}(x)\approx &\frac{3}{5}\ln\frac{1}{x}\,,\quad F_{1, \pm 1}^{\pm 1}(x)\approx F_{1, \pm 1}^{\mp 1}(x) \approx \frac{4}{5}\ln\frac{1}{x}\,.  
\gal
while at large $x$:
\ball{app7}
F_{10}^{\pm1}(x)\approx \frac{2\zeta(3)}{7x^2}\,,\quad F_{1, \pm 1}^{\pm 1}(x)\approx\frac{\pi^4}{3x^3}\,,\quad F_{1, \pm 1}^{\mp 1}(x)\approx\frac{\pi^2}{2x}\,.
\gal


\section{Chiral contributions to the cross section.}\label{appB}

Since in the ultrarelativistic limit the right (left) quark radiates the right (left) intermediate photon one can compute the chiral contributions to the cross section by considering the right and left-handed photon polarizations instead of summing over them in \eq{d55}:
\bal
&\left|\b\epsilon_{\b q'R}\cdot\left(\b T^{h*}_{k'lm}( \b q')+ih \b P^{h*}_{k'lm}(\b q') \right)\right|^2= \frac{1}{2}\left| ( \unit \xi+ i\unit \eta) \cdot \left(\b T^{h*}_{k'lm}(\b q')+ih \b P^{h*}_{k'lm}(\b q') \right)\right|^2\nonumber\\
=&\frac{1}{2}(1+h)^2\left| \unit \xi\cdot \b T^{h*}_{k'lm}( \b q')+ i \unit \eta\cdot \b T^{h*}_{k'lm}( \b q')\right|^2\,,\label{appB1}\\
&\left|\b\epsilon_{\b q'L}\cdot\left(\b T^{h*}_{k'lm}( \b q')+ih \b P^{h*}_{k'lm}(\b q') \right)\right|^2 
=\frac{1}{2}(1-h)^2\left| \unit \xi\cdot \b T^{h*}_{k'lm}( \b q')- i \unit \eta\cdot \b T^{h*}_{k'lm}( \b q')\right|^2\,.\label{appB2}
\gal
Thus, $h=\mp 1$ does not contribute to the right (left) polarization. 

To verify that 
the sum of  \eq{appB1} and \eq{appB2} yields \eq{d55} one can write it as 
\bal
&\frac{1}{2}(1+h)^2\left| \unit \xi\cdot \b T^{h*}_{k'lm}( \b q')+ i \unit \eta\cdot \b T^{h*}_{k'lm}( \b q')\right|^2+\frac{1}{2}(1-h)^2\left| \unit \xi\cdot \b T^{h*}_{k'lm}( \b q')- i \unit \eta\cdot \b T^{h*}_{k'lm}( \b q')\right|^2\nonumber\\
=& 2 \left| \unit \xi\cdot \b T^{h*}_{k'lm}( \b q')\right|^2 + 2\left| \unit \eta\cdot \b T^{h*}_{k'lm}( \b q')\right|^2 - 4h\im \left( \unit \xi\cdot \b T^{h*}_{k'lm}( \b q')\unit \eta\cdot \b T^{h*}_{k'lm}( \b q')\right)\,.
\label{appB4}
\gal
On the other hand, \eq{d55} can be written as 
\bal
2\left|  \unit \xi \cdot \left(\b T^{h*}_{k'lm}(\b q')+ih \b P^{h*}_{k'lm}(\b q') \right)\right|^2=
2\left| \unit \xi\cdot \b T^{h*}_{k'lm}( \b q')+ ih \unit \eta\cdot \b T^{h*}_{k'lm}( \b q')\right|^2\,,
\gal
which equals \eq{appB4}. Thus, the chiral contributions to the  cross section \eq{d73} are
\ball{appB6}
\frac{d\sigma_{lm}^h(R)}{dk'}= \frac{d\sigma_{lm}^{+1}}{dk'}\,,\qquad \frac{d\sigma_{lm}^h(L)}{dk'}= \frac{d\sigma_{lm}^{-1}}{dk'}\,.
\gal


\end{document}